\documentclass[runningheads]{llncs}
\usepackage{graphicx}
\usepackage{xcolor}
\usepackage[utf8]{inputenc}
\usepackage{fancyvrb,fvextra}
\usepackage{enumitem}
\usepackage{xurl}
\usepackage[colorlinks, citecolor=blue]{hyperref}

\begin{document}
\title{Load Balancer Tuning: Comparative Analysis of HAProxy Load Balancing Methods}
\author{Connor Rawls, Mohsen Amini Salehi}
\authorrunning{C. Rawls et al.}
\institute{High Performance Cloud Computing (\href{http://hpcclab.org/}{HPCC}) Lab,\\ University of Louisiana, Lafayette LA 70508, USA}
\maketitle              
%
\begin{abstract} Load balancing is prevalent in practical application (e.g., web) deployments seen today. One such load balancer, HAProxy, remains relevant as an open-source, easy-to-use system. In the context of web systems, the load balancer tier possesses significant influence over system performance and the incurred cost, which is decisive for cloud-based deployments. Therefore, it is imperative to properly tune the load balancer configuration and get the most performance out of the existing resources. In this technical report, we first introduce the HAProxy architecture and its load balancing methods. Then, we discuss fine-tuning parameters within this load balancer and examine their performances in face of various workload intensities. Our evaluation encompasses various types of web requests and homogeneous and heterogeneous back-ends. Lastly, based on the findings of this study, we present a set of best practices to optimally configure HAProxy. 
\end{abstract}

\section{Introduction}
\subsection{Load Balancing}
In a web system, the load balancer is usually the first component of the architecture to interact with incoming user requests. Load balancers are used to distribute user requests to application servers that can compute and generate the response. Two major goals of the load balancer in a web system are: A) To maximize the overall resource utilization; and B) To minimize the time it takes for each user to receive a response to their request. These two goals directly affect the QoS of a web application. To attain these goals, the load balancer uses \emph{load balancing methods} (algorithm) to distribute the user requests to compute machines. One study conducted by Google suggests that after a request's first 3 initial seconds, the probability that a user will leave the web application is 32\% \cite{dropRate}. In the case where the compute servers simply do not have enough resources to handle the incoming load of requests, the system should be scaled out to accommodate these load surges such that the application QoS is not violated. 

Cloud providers \cite{awsScaling}, \cite{azureScaling} offer services to automatically scale system resources under developer-defined conditions, such as when the CPU utilization threshold of an application server is met. The scaling of resources comes with a penalty, however. Increasing system resources for larger compute power incurs a higher cost to the user, and consumes more energy from the cloud providers perspective. Such consequences can be remarkable in a long run \cite{ica3pp10,li2016vlsc}. In the realm of bursty load behavior \cite{ipdps19,liperformanceanalysis}, applications may see large, unexpected spikes in operating costs if the acquisition of resources is allowed to run rampant. Therefore, there lies a delicate balance between maximizing the performance of the application while minimizing its cloud operating costs.

One method of maximizing system performance may lie within tuning various components of the web tier. Another method can be tuning the behavior of load balancer tier via its configurations. Most software may perform subpar or inadequately out-of-the-box. Users, however, can adjust certain settings either prior or during runtime to increase performance. Maximizing each component's performance may prove critical to some in decreasing the need for scaling out system resources, in turn, maintaining low operational expenditures \cite{denninnart2018leveraging,mahmood17}.

\subsection{HAProxy}
The idea of using load balancer in distributed systems has a long history and has been studied in various contexts, such as those for Grid computing \cite{Mohsen2006,mlbm06} and cloud \cite{li2018cost} or for different applications \cite{haproxy}. All these load balancers can be broadly categorized as Network Load Balancer (NLB) and Application Load Balancer (ALB). 
The load balancing software used in this paper's study is HAProxy. HAProxy is an open-source load balancer meant to be as stateless as possible while maintaining high throughput of messages per second. Additionally, HAProxy can be configured to be used in both Network Load Balancing and Application Load Balancing contexts.

\begin{figure}
  \begin{center}
  \includegraphics[width=3.5in]{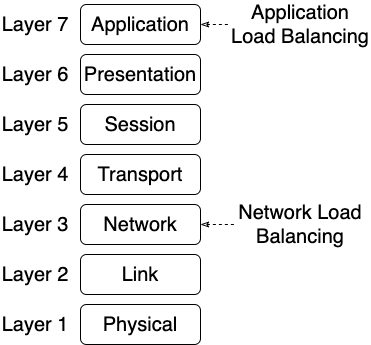}\\
  \caption{Overview of Open Systems Interconnection model.}\label{osi}
  \end{center}
\end{figure}

\paragraph*{Network Load Balancer (NLB):} NLBs describe load balancers that operate solely at the transport level (i.e., TCP). More specifically, NLBs operate at the Open Systems Interconnection (OSI) Layer 4, highlighted in Fig.~\ref{osi}. This figure highlights the separate levels of information contained in the common communication packet exchanged between two agents. Network load balancing is not concerned with the intricacies of the messages it is handling, such as their content, headers, etc. Furthermore, NLBs do not consider the behavior of backend servers in its decision making. Instead, NLBs only consider transport-related information when routing messages and that is why they are often faster than the ALB algorithms.

\paragraph*{Application Load Balancer (ALB):} Opposing NLBs, ALBs do consider the contents of the messages it is routing when making decisions. From a technical standpoint, this means that ALBs operate at OSI Layer 7 (ex. HTTP), as shown in Fig.~\ref{osi}. In HTTP scenarios, this means that an ALB routing algorithm may consider application layer fields such as the request method or the URL requested. For example, an ALB might route requests whose messages contain certain header values to one specific backend server. Additionally, ALBs can potentially include the state information of backend servers in their logic, hence, are often more efficient than their NLB counterparts.
\\

 In terms of HAProxy, the mechanism for enabling either NLB or ALB routing is dependent on the load balancing algorithm chosen before runtime. HAProxy comes with several builtin load balancing algorithms that are commonly used in production, such as round-robin, least-connection-based, and random. It is important for the solution architects to choose the appropriate load balancing algorithm (method) based on the characteristics of the system they are deploying, as each possesses different behaviors depending on the workload. The algorithms that HAProxy provides can be contextual in either an ALB or NLB scenario. For example, HAProxy supports URL hashing to ensure that specific paths on one's website is always directed to the same server(s). This algorithm can be considered ALB-based, as the logic examines the intricacies of the message itself as well as the state of the backend servers to make decisions. On the other hand, the traditional round-robin algorithm that HAProxy provides chooses backend servers in order, hence, is considered as NLB-based.

 Seeing the complexity and prevalence of load balancing and HAProxy, the purpose of this paper is to take a deep dive into HAProxy and provide insight on its inner workings. In a production environment, a minor improvement in HAProxy can have substantial impacts on the user satisfaction and the incurred cost of deploying an application. 
 
 In summary, the contributions of this work are as follows:
 \begin{itemize}[label=\textbullet]
  \item Explanation of load balancing methods and HAProxy's implementation.
  \item Exploration of tuning HAProxy configuration parameters.
  \item Performance comparison of different load balancing algorithms under various workload scenarios.
  \item Impact of server heterogeneity and homogeneity on the performance of various load balancing algorithms.
\end{itemize}

\begin{figure}
  \begin{center}
  \includegraphics[width=\linewidth]{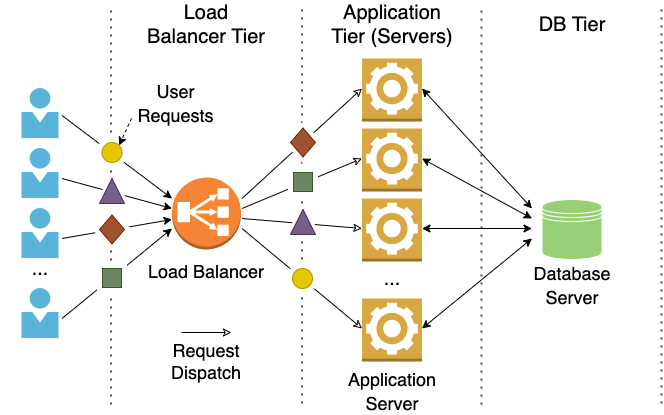}\\
  \caption{Architecture of general three-tier web system. The different shapes passing into and out of the load balancer represent different task types.}\label{systemScope}
  \end{center}
\end{figure}

\section{Load Balancing Architecture}
In web application deployments, the system can be broken down into three separate tiers: load balancing, application, and database. An overview of such an architecture can be observed in Fig.~\ref{systemScope}. The load balancing tier accepts incoming user requests. With the load balancing algorithm, an appropriate back-end server is determined from a list of possible servers and the request is then dispatched to the application. The application tier's purpose is to satisfy the computational workload that the request brings. The application machines rely on data that is present in the database tier. With incoming requests, the application tier queries the database tier for information used to handle its workload. Once a request has completed its execution in the application tier, the server sends a response to the load balancer to ultimately be returned to the client, completing the transaction.

Web requests are often user-facing, in that there is some deadline the response must meet. The deadline can be considered a concept developed between the inter-client/company relation. It has been shown that as the web response time grows, the satisfaction of clients begins to drop linearly \cite{hoxmeier2000system}. In the circumstance of website hosting, this may result in a loss of traffic and, in turn, company profits. Other situations in which communication is considered mission critical, such as in healthcare environments, slow response times may result in catastrophic failure. Therefore, there is a call for ensuring the efficient load balancing of requests to minimize response times and make sure that the requests are served within that time.

\section{HAProxy Architecture Unfolded}
\begin{figure}[h]
  \begin{center}
  \includegraphics[width=\linewidth]{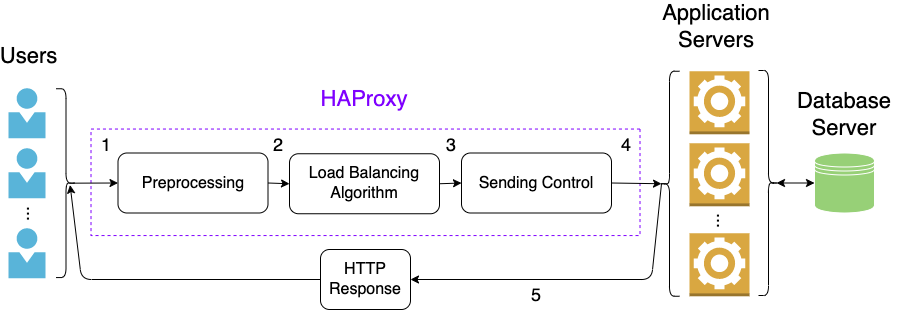}\\
  \caption{HAProxy internals, generalizing the main mechanisms used to load balance incoming user requests to application servers.}\label{hapArch}
  \end{center}
\end{figure}

The internal architecture of HAProxy can be viewed in Fig.~\ref{hapArch}. In step 1, incoming user requests are received by HAProxy. HAProxy performs various preprocessing functions on these messages such as determining header values, paths requested, and networking information. In Step 2, the request is passed to the load balancing algorithm. This step is where HAProxy determines which server to dispatch the user request. In Step 3, HAProxy handles the request-to-server task in a series of queued tasks. HAProxy establishes network sockets necessary to deliver the message to the assigned server in Step 4. Lastly, Step 5 highlights the application server's finalized response to the user request.

\section{Load Balancing Algorithms Supported by HAProxy}
The algorithm used in load balancing is critical to the behavior of the load balancer itself. Each algorithm may exhibit significantly different performances in terms of response times and error rates. Additionally, choosing the correct algorithm may prove difficult for non-technical users. As such, it is important to have a general understanding of in which scenarios to use particular algorithms.

\begin{figure}
  \begin{center}
  \includegraphics[width=\linewidth]{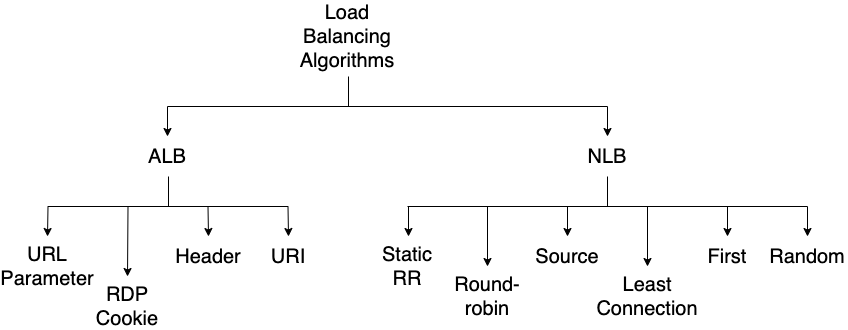}\\
  \caption{All algorithms supported by HAProxy, divided into NLB-based and ALB-based logic.}\label{algs}
  \end{center}
\end{figure}

The load balancing algorithms supported by HAProxy can fall into two categories: ALB or NLB. This categorization can be viewed in Fig.~\ref{algs}. We will discuss each algorithm in due order but first, we will make note of an important mechanism that HAProxy utilizes in a few of its algorithms.

Under some scenarios, such as in heterogeneous environments/workloads, certain application servers may wish to be prioritized or considered more heavily in load balancing decisions. As such, HAProxy makes use of a weighting mechanic for each server. What this tool provides is a way for the load balancing algorithm to make conditional decisions based upon a server's priority in relation to the other servers. For example, a server more heavily weighted generally signifies that the algorithm prefers to dispatch requests to this particular server in comparison to other servers that are not weighted as high. By default, HAProxy sets the weight of all servers to the same static value of 1.

\subsection{Random}
Random, otherwise known as Power of Two \cite{richa2001power}, randomly pulls two servers from the list of possible servers. From these two servers, the algorithm chooses the server with the least current load (connections). This algorithm can further be adjusted to support Power of N where N is any positive integer. One can expect that as N approaches the number of actual backend servers, the algorithm's performance will begin to mimic the least connection-based algorithm.

\subsection{First}
With the \textit{first} algorithm, an incoming request is dispatched to the first server possessing an available connection slot. The selection of possible backend servers are treated as a list or pool. This list is sorted based upon each server's id, which is some value designated by a system administrator. Upon a routing decision, HAProxy will select the previously used server from the list (in the case of the first request during the application's startup, this is the first server in the list). HAProxy will continue to route incoming requests to this same server until the server's designated max connection value is reached. From here, HAProxy will then send requests to the server that is next in line.

This algorithm may prove useful for utilizing the smallest amount of servers possible, maintaining low operational costs. However, in times in which a server is approaching its max connection value, the tasks in execution are likely to suffer. This is due to the numerous tasks competing for system resources.

\subsection{Least Connection}
HAProxy's least connection algorithm is based upon the connection state of each server. In dispatching an incoming request $\textit{r}$, out of the list of possible servers, $\textit{r}$ will be routed to the server with the current least number of connections. If two or more servers possess the same number of connections, the round-robin algorithm is used to determine between this subgroup.

\subsection{Source}
Under the source algorithm, the client's IP of the incoming request is hashed utilizing the sum weight of all of the running servers. Utilizing this hash, HAProxy will dispatch the request accordingly. The consideration of the sum weight means that future requests from the same clients will always be routed to the same servers. However, these mappings would change given that a server joins or leaves the backend. Consequently, most clients would then be routed to a separate server.

\subsection{Round-robin}
With the traditional round-robin algorithm, a selection cycle is used to iterated through the list of possible servers. Each server will be selected at most once per selection cycle. HAProxy utilizes a weighted version of the round-robin algorithm. Weighted round-robin functions similarly in that there is a selection cycle for routing, however, each server is assigned a weight to be considered in the algorithm's logic. The weight of each server is used to determine the proportion of requests to be dispatched to it. For servers \emph{m} and \emph{n} in the selection pool with weights \emph{i} and \emph{j} respectively and \emph{i} $>$ \emph{j}, the first \emph{i} incoming requests will be dispatched to server \emph{m}. Upon the \emph{i} $+$ 1 request, the next \emph{j} requests will be dispatched to server \emph{n}.

The weighted round-robin algorithm is useful for heterogeneous systems or systems in which the servers are susceptible to dynamic performance changes. However, the difficult part arises in that the system administrator must understand what the weight of each server should be prior to operation. This manual input invites human error into the load balancing equation.

\subsection{Static RR}
The Static RR (Round-Robin) algorithm in HAProxy functions similarly to its non-static counterpart. This difference between these two, however, lies in the fact that Static RR does not support changing a server's weight on the fly. A server's weight may be altered during HAProxy's runtime as the system admin deems fit. This procedure is useful for manually tweaking the load balancing logic, providing a more granular management. Some algorithms, such as Static RR, do not take advantage of this feature. Instead, the server weights are observed during startup and considered during the rest of the algorithm's lifespan.

One might consider utilizing this algorithm in scenarios where CPU resource expenditure is very tight. This is due to the algorithm's slightly less CPU intensive nature as compared to RR.

\subsection{URI}
HAProxy's URI algorithm is a statically hashed routing method, similar to \textit{source}. In URI, a selective part or the entirety of each request's URI is passed through the hashing algorithm. The output value is modified using the sum of the backend server weights, providing similar properties to \textit{source}. Users of this algorithm can decide to utilize either the query, the selected path, or both of these parameters of the request for the hashing procedure. In addition to these customization options, URI also supports the \textit{depth} parameter. This value controls the how far into a directory a request's path is for the algorithm's consideration.

This algorithm could prove useful for ensuring that all requests pertaining to a particular page, or all pages beyond a certain depth within a directory, are routed to the same server. Accordingly, pages that are known to be heavily trafficked or possess an inherently large computational load can be allocated to servers that contain proportionally larger hardware resources. In contrast, requests that are non-intensive can be directed to "smaller" servers.

\subsection{Header}
The header algorithm is another hashing method that uses a user-specified header value to be hashed in determining the server to be routed to.

\subsection{RDP Cookie}
The RDP Cookie algorithm examines the name in the RDP cookie for its load balancing decisions. This value is hashed and assigned to a corresponding server. This method ensures that returning clients will continuously be assigned to the same server.

\subsection{URL Parameter}
With the URL Parameter algorithm, the query string of each request is used for the hashing algorithm. If no query was found in the request, this algorithm resorts to the round-robin algorithm.

This algorithm may prove useful for ensuring that returning clients will be routed to the same server, given that a server has not left or entered the pool of possible servers since their last request.

\section{Fine-Tuning HAProxy}
While utilizing HAProxy out-of-the-box may prove applicable under certain scenarios, it is important to maximize the performance of load balancers under most cases. In the context of resource scaling, additional resources will be initiated or spawned upon meeting some insufficient performance metric such as excessive response times of requests or high CPU utilization of compute servers. This scaling increases performance in times of increased load but comes with an increased operational costs. To minimize the need to scale out, maximizing the efficiency of your current system is imperative.

\subsection{Configuration Tuning}
Being that HAProxy possesses over 50 possible parameters to configure, we select only the tuning parameters that are influential in performance for most load balancing environments. These parameters can be observed in Table~\ref{tunparam}.

\begin{table}
\begin{center}
    \begin{tabular}[ht!]{ |p{2cm}|p{5cm}| }
     \hline
     \textbf{Parameter} & \textbf{Note}\\ [0.5ex]
     \hline\hline
     nbproc & Number of processes\\
     \hline
     nbthread & Number of processing threads\\
     \hline
     cpu-map & Designate specific CPU cores for specific threads to process on\\
     \hline
     maxconn & Maximum number of concurrent connections HAProxy will allow\\
     \hline
     busy-polling & Prevents processor from sleeping during idle periods\\
     \hline
     compression & Compresses HTTP messages\\
     \hline
     spread-checks & Spread out health checks to servers instead of sending all at once\\
     \hline
    \end{tabular}
    \vspace*{5mm}
    \caption{Selected tuning parameters for increasing HAProxy performance.}\label{tunparam}
    \end{center}
\end{table}

The max connection parameter of HAProxy is a network configuration option that allows users to control how many potential client connections can be established at one time. Given a value of \textit{x} that is presented to the \textit{maxconn} parameter, HAProxy will reject incoming connections if there are currently \textit{x} connections already established. Users can set this value as high as their load balancing server's ulimit value will allow. Setting this value below the server's ulimit could prove useful in securing memory resources to background and OS processes. Additionally, limiting the max connection value may prove integral for security reasons.

HAProxy's \textit{nbproc} directive can be used to spawn more HAProxy processes. Each one of these processes will handle a portion of the overall HAProxy computational load. In addition, \textit{nbthread} can be used to further parallelize HAProxy. It would seem obvious that distributing the workload of HAProxy and executing each simultaneously can significantly decrease the overall computational time required to load balance client requests. However, there are certain drawbacks that should be considered upon implementing these forms of parallelization. For one, the \textit{nbproc} directive does not support data sharing between processes. To combat this issue, \textit{nbthread} could be used in lieu. Additionally, HAProxy uses health checks to obtain state information on the backend servers. This means that a dummy request is periodically sent to the backend servers. With \textit{nbproc}, each process will send its own health checks, resulting in increased network traffic. Lastly, increasing the thread count beyond reason is detrimental to performance. If the number of threads in execution represent a pool of workers that exist in an environment that can not adequately provide enough CPU time, the resulting contention (CPU thrashing) will lead to each thread possessing less time to compute. Therefore, there exists a balance for allocating the proper number of processes/threads.

To have an even more granular control on HAProxy's processing, one can utilize \textit{cpu-map}. This directive allows users to control which process executes on which CPU core. Essentially, the designated process(es) will always execute on the designated CPU core.


There exists many other such tuning parameters such as \textit{tune.bufsize}, which alters the amount of memory each process is allocated or \textit{nosplice}, which disables the kernel's ability to perform TCP socket splicing. However, HAProxy's documentation suggests that enabling/changing these parameters may cause buggy behavior or even result in the communication of corrupted data. As such, HAProxy also recommends that these parameters not be touched outside of their own core development team or under very specific scenarios. As our experiments are meant to remain representative of common use-case environments, these specialized parameters will not be explored in this work.

\section{Performance Evaluation of HAProxy}

\subsection{System Setup}
In this study, a three-tier web system was created to replicate a realistic environment that users might find while browsing a website. A blog and ecommerce website was created and propagated with sample pages, items, and posts. This website is used as the application in our architecture. Two types of web requests (task types) were generated and profiled, each containing 5 possible instance types, as explained in the next sections. The metrics were captured utilizing 40,000 instances of each task type executed in isolation to ensure no interference from outside resource contention.

\begin{figure}
  \begin{center}
  \includegraphics[width=\linewidth]{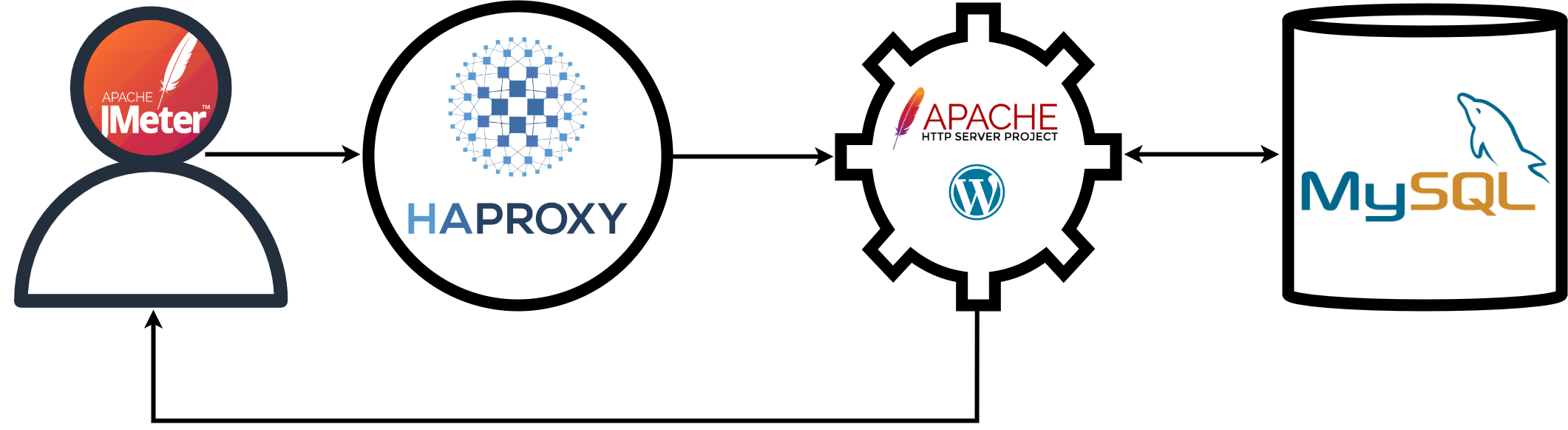}\\
  \caption{Design of our experimental system \cite{jmeterIcon}, \cite{hapIcon}, \cite{apacheIcon}, \cite{wpIcon}, \cite{mysqlIcon}.}\label{expArch}
  \end{center}
\end{figure}

A view into the hardware and software specifications of our experimental system can be found in Table~\ref{hardsoft}. The web service architecture was established in a mixture of bare-metal, virtual, and containerized machines. The user requests were synthesized using Apache Jmeter \cite{jmeter} from possible requests on the website. The user server sits in its own virtual machine. This removes memory contention from other working machinations of the system and provides a realistic representation of network messaging. The load balancer and application servers each reside in their own container. The application machines are all completely identical in terms of hardware and software. The database is a separate bare-metal machine from the one that the user, load balancer, and application machines reside on. Fig.~\ref{expArch} showcases our specific software environment for each tier.

\begin{table}
\begin{center}
    \begin{tabular}[ht!]{ |p{2.45cm}|p{1.25cm}|p{4cm}|p{2cm}| }
     \hline
     \textbf{Tier} & \textbf{Server Count} & \textbf{Underlying Hardware} & \textbf{Software}\\ [0.5ex]
     \hline\hline
     User & 1 & 8~cores~(2.30~GHz) / 32 GB & KVM VM\newline
     with Apache\newline JMeter\\
     \hline
     Load Balancer & 1 & Container on Host with 112~cores~(2.20~GHz) / 330 GB & Docker Container with\newline HAProxy\\
     \hline
     Application & 5 & 16~cores~(1.80~GHz) / 32 GB & KVM VM\newline with Apache\newline and WordPress\\
     \hline
     Database & 1 & 48~cores~(2.87~GHz) / 132 GB & Bare-metal with MySQL\\
     \hline
    \end{tabular}
    \vspace*{5mm}
    \caption{Hardware and Software Specifications Used for Experimentation}\label{hardsoft}
    \end{center}
\end{table}

\subsection{Input Workload}
The examined website created through WordPress \cite{wordpress} with various community-sourced plugins. This allowed us to propagate the website with multiple pages and contents.

Jmeter was used to record the possible HTTP requests that could be sent to the website. These requests are saved as XML files to be used for profiling later. Each file can be considered as a single test scenario. Each scenario is composed of only two task types: GET and POST. In the context of HTTP web requests, GET tasks are used by users to fetch information from the web service, such as a web page or image. POST tasks are used by users to send information to the web service, such as posting a comment or image to the web page. Being that a long sequence of POST requests would cause our web pages to become bloated or change over time, consequentially, sequential requests would be affected. Therefore, we have implemented an instrument to immediately remove POST request content from the web page as soon as they are processed, leaving the web page unchanged from the time of its initial creation.

The scenarios used for testing vary in the request rate that is provided to the system. The first scenario's request rate was approximately 16.67 users/second over a period of 60 seconds for a total of 1,000 requests. We then ran similar tests with differing request rates of over the same period providing a list of total requests: [5,000:40,000] with increments of 5,000 total requests. The total amount of requests can be partitioned between both task types, however, requests instances of both task types were sent simultaneously. The input workload scenario configurations can be observed in Table~\ref{workscen}.

\begin{table}
\begin{center}
    \begin{tabular}[ht!]{ |p{3cm}|p{3cm}|p{3cm}| }
     \hline
     \textbf{Total Requests} & \textbf{Test Period (s)} & \textbf{Requests/Second}\\ [0.5ex]
     \hline\hline
     1000 & 60 & 16.67\\
     \hline
     5000 & 60 & 83.33\\
     \hline
     10000 & 60 & 166.67\\
     \hline
     15000 & 60 & 250\\
     \hline
     20000 & 60 & 333.33\\
     \hline
     25000 & 60 & 416.67\\
     \hline
     30000 & 60 & 500\\
     \hline
     35000 & 60 & 583.33\\
     \hline
     40000 & 60 & 666.67\\
     \hline
    \end{tabular}
    \vspace*{5mm}
    \caption{Input Workload Scenarios for Testing HAProxy Performance.}\label{workscen}
    \end{center}
\end{table}


\begin{figure}
  \begin{center}
  \includegraphics[width=10cm]{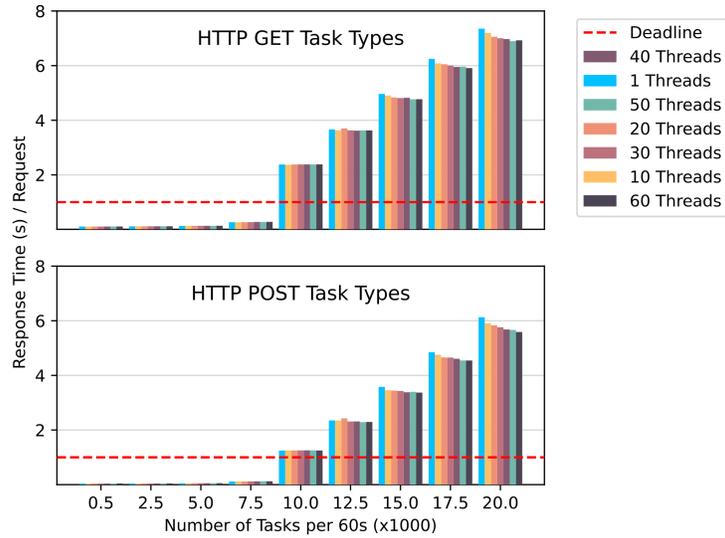}\\
  \caption{Performance results from tuning processing threads parameter. The dashed red line indicates the deadline tasks should be meeting.}\label{threadTuning}
  \end{center}
\end{figure}

\section{Tuning Number of HAProxy Threads}
As discussed earlier, getting the most out of our current load balancer is useful for decreasing system costs and maximizing system performance. For this experiment, we measure the performance of various configurations of HAProxy. Being that HAProxy processes many possible parameters, each with a large amount of possible parameter values, the combinatorial search space to examine all configurations is time prohibitive. Hence, we examine the number of processing threads parameter as a potentially influential parameter on the performance. We utilize the round-robin load balancing algorithm, as this is the most general algorithm provided by HAProxy and can be utilized in most non-specialized scenarios. The results of this test can be observed in Fig.~\ref{threadTuning}.

It could be expected that increasing the number of processing threads leads to an increase of the number of tasks executed in parallel. Hence, we should expect that the response times of requests to decrease as the load balancer is able to accommodate incoming tasks accordingly. As our experiment shows, the number of threads in our case marginally affects the response time. A possible explanation as to why we do not see any significant effect that the number of threads possess on response time is that our backend system is not complex enough to exploit this feature of HAProxy. Our system is small enough so as to allow HAProxy to already minimize the response time of requests with just one thread. Essentially, in our case, we receive no use out of increasing the parallelism of HAProxy because there is nothing to be parallelized.

From our results, we can form a safe estimation that the default number of processing threads provided by HAProxy (64) is enough  to accommodate most workloads. Furthermore, decreasing the number of threads controlled by HAProxy will only marginally decrease the response time of requests. As such, setting nbthread to 1 will allow for similar performances while alleviating the potential contention for resources for additional applications running on the same machine as HAProxy.\\
\\
\noindent\fbox{%
\parbox{\textwidth}{
\textbf{Best Practice} Reduce the number of processing threads to 1 if HAProxy is sharing system resources with additional processes. Otherwise, it is best to leave nbthread untouched.
}%
}

\section{Comparing Performance Impact of Load Balancing Algorithms}
\begin{figure}
  \begin{center}
  \includegraphics[width=9.9cm]{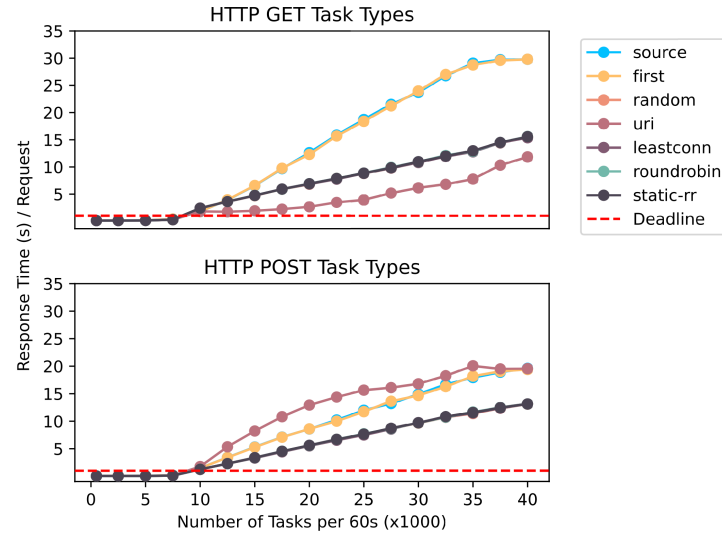}\\
  \caption{Performance results from tuning the load balancing algorithm. Due to trend overlaps, the following algorithms are stacked: [fist, source], [random, leastconn, static-rr, roundrobin], [uri].}\label{algorithmTuning}
  \end{center}
\end{figure}

It is important to understand when to use a specific load balancing algorithm for the task at hand. If the incoming workload's characteristics are known prior to operation, their behavior can be exploited to choose the correct algorithm in order to maximize system performance. In this experiment, we compare each algorithm against the same workloads as in the previous experiment. The algorithm performance results can be viewed in Fig.~\ref{algorithmTuning}. It should be noted that some algorithms supported by HAProxy require a preparation period to designate the algorithm logic. For example, the RDP cookie algorithm assumes remote desktop operation in the system environment. As our scenarios do not partake in such action, we have obligated to remove these algorithms for the following experiment. Additionally, it was originally observed that the algorithms experiment begin to show interesting results at the 40,000 task count mark. Due to this behavior, we increased the range of tasks to 80,000 to better view a discrepancy between the algorithms' performance.

From the performance results, we can observe that most algorithms appear to be impartial to the task types in our given workload. URI, however, exhibits different performance for each task type. For example, URI shows the best performance for GET requests while exhibiting the worst performance for POST requests. The likely cause for this behavior is due to the URI algorithm directing all GET requests to the same servers while sending POST requests to others. In this scenario, URI is making a task-to-server partitioning scheme that is resulting in a loss of overall performance. Alternatively, the other algorithms send a mixture of both task types to all servers. It can be speculated that if our environment consisted of a set of servers to better accommodate the resources necessary to respond to the POST requests, the URI algorithm would showcase the best performance or both task types. However, being that our backend is homogeneous, the potential for URI's logic can not be fully utilized. Additional to this notable feature, the first and source algorithms seem to perform worse for GET tasks. Comparatively, the static-rr, random, roundrobin, and leastconn algorithms are more robust to task types as these algorithms do not exhibit much of a performance change as the number of requests increase between task types.

From the results of the load balancing algorithm experiment, it is prevalent that some ALB algorithms may require further system tuning to receive the full benefits they bring in terms of request response time. However, while choosing an NLB might appear to be a safer, more general alternative, there are still significant performance discrepancies between them.\\
\\
\noindent\fbox{%
\parbox{\textwidth}{
\textbf{Best Practice} Choose random, leastconn, roundrobin, or static-rr for load balancing under general circumstances. Choose URI for potential performance increase at the cost of system profiling/further resource tuning.
}%
}

\section{Impact of Load Balancing Algorithms in a Heterogeneous Environment}
\begin{figure}
  \begin{center}
  \includegraphics[width=9.9cm]{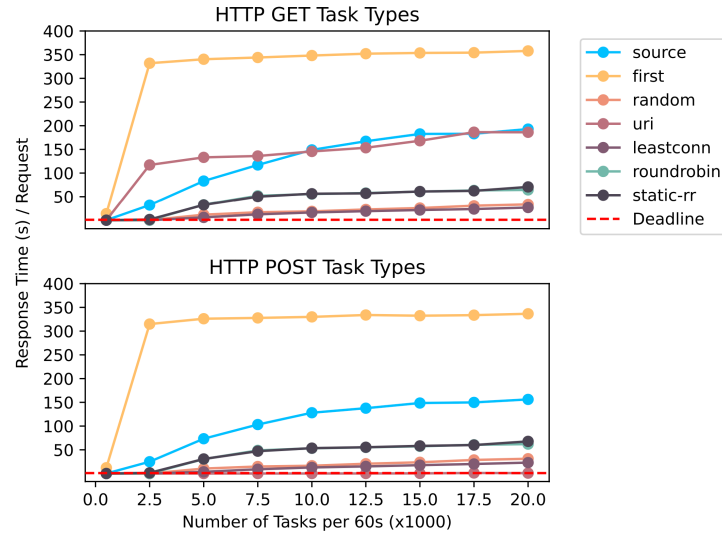}\\
  \caption{Performance results of heterogeneous environment experiment. Due to trend overlaps, the following algorithms are stacked: [static-rr, roundrobin], [random, leastconn].}\label{heterogenResults}
  \end{center}
\end{figure}

As we have seen from the load balancing algorithm experiment, there lies potential performance differences when the server environment is heterogeneous (URI results). Due to these observations, we have ran the same load balancing performance experiment on a different set of server configurations. The heterogeneous server environment hardware specifications can be found in Table~\ref{heteroHardware}. The hardware configurations used in this experiment were inspired by AWS's M5 EC2 instances. As such, our configurations are meant to be representable of these instance types. The results of this experiment can be observed in Fig.~\ref{heterogenResults}.

\begin{table}
\begin{center}
    \begin{tabular}[ht!]{ |p{2cm}|p{1.5cm}|p{4.5cm}| }
     \hline
     \textbf{EC2\newline Instance} & \textbf{Server Count} & \textbf{Underlying\newline Hardware}\\ [0.5ex]
     \hline\hline
     m5.xlarge & 1 & 4~cores~(1.80~GHz) / 16 GB\\
     \hline
     m5.2xlarge & 1 & 8~cores~(1.80~GHz) / 32 GB\\
     \hline
     m5.4xlarge & 1 & 16~cores~(1.80~GHz) / 64 GB\\
     \hline
     m5.8xlarge & 1 & 32~cores~(1.80~GHz) / 128 GB\\
     \hline
     m5.12xlarge & 1 & 48~cores~(1.80~GHz) / 192 GB\\
     \hline
    \end{tabular}
    \vspace*{5mm}
    \caption{Heterogeneous environment hardware specifications used for experimentation.}\label{heteroHardware}
    \end{center}
\end{table}

It is known that certain load balancing algorithms exploit the nature of heterogeneous systems while others rely on the homogeneity of servers to perform well. As with the impact of algorithms experiment, URI performs well for tasks of one type but poorly for tasks of another type. For POST tasks, URI is able to meet the deadline for all scenarios except for an incoming task rate of 17.5k tasks/60s. Additionally, with the heterogeneous environment, we see less overlaps in performance trends between tasks. For example, in the homogeneous results, random, leastconn, static-rr, and roundrobin all exhibited the same performance behaviors. However, in the heterogeneous results, random and leastconn share similar behaviors alone while static-rr and roundrobin share behaviors.

We can also observe that the overall response times increase drastically when utilizing a heterogeneous server environment as compared to the homogeneous counterpart. For example, random showcases a relatively small curve for both task types in a heterogeneous environment. In the homogeneous results, random showed similar agnostic behavior in regards to task type, but possessed a 63.75\% lower response time for an incoming task rate of 20k tasks / 60s.

Load balancing algorithms source and URI both use information from incoming tasks to partition workloads to application servers in a manner that could be exploited by server heterogeneity. However, from these results, it can be claimed that URI can potentially show better performance at the cost of reliability in terms of task type while source shows better robustness to task task at the cost of response time. One potential explanation for this dynamic may lie within the differences in task characteristics NLB and ALB algorithms observe. NLB algorithms only examine network-level information about a given task, such as the user's IP address. This kind of information is not particular to a task and fails to provide important features that describes an incoming workload. On the other hand, ALB algorithms examine more granular information about a task, such as the page requested. Under the guise of an ecommerce store, particular pages may be more object-rich. From this example, an ALB algorithm will be able to determine the "weight" of an incoming task and better act accordingly, as compared to NLB algorithms.\\
\\
\noindent\fbox{%
\parbox{\textwidth}{
\textbf{Best Practice} It is imperative to utilize algorithms that can properly exploit server heterogeneity if the backend is not homogeneous. ALB algorithms that fit into this category (URI) show better performance results.
}%
}

\section{Extending HAProxy with Custom Load Balancing Methods}
Utilizing the available options HAProxy provides may prove suitable for many general use cases. However, it may be desired to further tune HAProxy's performance, we design custom load balancing algorithms to cater to specific use cases. Being that HAProxy is open-source, we can easily dive deep into its source code to edit the way an algorithm behaves or add our own custom algorithm. In this section, we will describe an example scenario for customizing a specific algorithm while highlighting important aspects of HAProxy's code. We will be referring to files and directories that can be found on HAProxy's GitHub repository \cite{hapGit}.

In our example, we will be customizing the load balancing algorithm random. We would like to inject additional information into its logic to consider for load balancing incoming requests. The current method provides ample task dispersion so as to not ``overcrowd" one specific server. For the sake of illustration, in our example, we desire to further increase the granularity of the equality of resource partitioning.

The load balancing algorithm that is desired to be edited is called from the file \verb|backend.c| in HAProxy's \verb|src/| directory. The \verb|assign_server| function is used to parse which algorithm the user has decided from their configuration file and call the algorithm itself. Additional to this, the HTTP request is passed as an object possessing various characteristics such as URL paths and headers. In our case, the \verb|get_server_rnd| function is called. Located in this function is the core of the load balancing logic.

For our example, we capture CPU utilization information from our application servers using the libvirt library \cite{libvirt}. We implement a custom library to capture this information remotely and embed it into HAProxy's code. The current CPU utilization of each server is recorded in real-time and called from \verb|get_server_rnd| before making its final decisions. Specifically, the random algorithm will only pull from a list of possible servers whose CPU utilization lies below a user-specified threshold. From here, we let the default random algorithm take over and return the selected server.

When adding custom libraries, it is necessary to include them in the \verb|Makefile|. Specifically, the \verb|OBJS| variable must include the new object file that is desired to be created from the new header file.

HAProxy's code is complex and should be respected when adding additional content. To avoid any potentially unwanted behaviors or errors, leaving HAProxy's code as default as possible is a good measure. In our example, we segregate our custom logic as much as possible from HAProxy's original logic. We attempt to minimize the association that our code holds with HAProxy and let it perform its own actions where it can. Customizing with this idea in mind will prevent any unforeseen consequences should an unassuming variable or block of code is interacted with, causing a snowball effect somewhere else in HAProxy's dense architecture. 

Once the code has been altered and it is time to compile, there is a list of options to choose from such as lua version, multithreading support, and target compiler. A list of these options can be found in the \verb|Makefile|. In our example case, we include the additional libvirt library to support the CPU utilization readings. To compile additional libraries that HAProxy would otherwise be unfamiliar with, the compile command must include the options \verb|ADDINC| and \verb|ADDLIB| for acknowledging the library path and adding the library to the list of libraries to be compiled, respectively. Here is our full command to compile with the included libvirt library:
\begin{Verbatim}[breaklines]
make -j \$(nproc) TARGET=linux-glibc USE\textunderscore LUA=1 LUA\textunderscore INC=/opt/lua-5.4.3/src/ LUA\textunderscore LIB=/opt/lua-5.4.3/src/ ADDINC=-L/opt/libvirt ADDLIB=-lvirt
\end{Verbatim}
From here, one can simply \verb|make install| and then run the customized HAProxy program.

\section{Conclusion and Future Works}
In this work, we studied HAProxy load balancer. We explored its internal mechanics and its parameters. Specifically, we unfolded the load balancing algorithms it uses and examined them under real-world settings with various workload intensities and under homogeneous and heterogeneous back-end servers. We realized that: (A) When running HAProxy in isolation on a given server, using the default value of processing threads leads to the best performance in terms of response time. If HAProxy is sharing the host machine with additional processes, then users might be inclined to lower the number of threads HAProxy controls. Our experiments show that this will come with a marginal increase in the response time of requests. (B) Under general load balancing scenarios in a homogeneous server environment, the algorithms that provide the lowest response times are random, leastconn, roundrobin, and static-rr. Furthermore, these algorithms exhibit little deviation amongst themselves in trends in response time. Utilizing algorithms that can exploit server heterogeneity may lead to potential increases in performance at the cost of necessary system profiling and further resource tuning. (C) If using a heterogeneous server environment, it is applicable to use algorithms that perform well under homogeneous scenarios. Therefore, it is necessary to characterize the servers and implement either source or URI algorithms to receive the acceptable performance for the deployed application. (D) ALB algorithms that exploit heterogeneous natures show better performance than their NLB counterpart.

In the future, we plan to extend HAProxy with custom-designed load balancing algorithms. Importantly, we plan to implement the ability to proactively drop task requests that are unlikely to meet their deadlines. Such a method can have at least two benefits: (A) coping with the oversubscribed situation and make the system busy for tasks with a higher chance of meeting their deadlines; and (B) handling Denial of Service (DoS) attacks and proactive drop (ignore) tasks that are artificially generated to make the system unresponsive.
\bibliographystyle{plain}
\bibliography{bib}
\end{document}